\documentclass[twocolumn,secnumarabic,amssymb, aps, prl, figure, superscriptaddress,floatfix, tightenlines] {revtex4-1}
\usepackage{graphicx}

\usepackage{braket}

\begin{document}

\title{A New Perspective on the Role of A-site Cation in Perovskite Solar Cells}%
\author{Chang Woo Myung}
\affiliation{Center for Superfunctional Materials, Department of Chemistry and Department of Physics, Ulsan National Institute of Science and Technology (UNIST), Ulsan 44919, Korea}

\author{Jeonghun Yun} 
\affiliation{Center for Superfunctional Materials, Department of Chemistry and Department of Physics, Ulsan National Institute of Science and Technology (UNIST), Ulsan 44919, Korea}

\author{Geunsik Lee}
\affiliation{Department of Chemistry and Department of Physics, Ulsan National Institute of Science and Technology (UNIST), Ulsan 44919, Korea}

\author{Kwang S. Kim}
\affiliation{Center for Superfunctional Materials, Department of Chemistry and Department of Physics, Ulsan National Institute of Science and Technology (UNIST), Ulsan 44919, Korea}

\date{\today}%


\begin{abstract}
As the race towards higher efficiency for inorganic/organic hybrid perovskite solar cells (PSCs) is becoming highly competitive, a design scheme to maximize carrier transport towards higher power efficiency has been urgently demanded. Here, we unravel a hidden role of A-site cation of PSCs in carrier transport which has been largely neglected, i.e., tuning the Fröhlich electron-phonon (e-ph) coupling of longitudinal optical (LO) phonon by A-site cations. The key for steering Fröhlich polaron is to control the interaction strength and the number of proton (or lithium) coordination to halide ion. The coordination to I$^-$ alleviates electron-phonon scattering by either decreasing the Born effective charge or absorbing the LO motion of I. This novel principle discloses lower electron-phonon coupling by several promising organic cations including hydroxyl-ammonium cation (NH$_3$OH$^+$) and possibly Li$^+$ solvating methylamine (Li$^+$\textperiodcentered \textperiodcentered \textperiodcentered NH$_2$CH$_3$) than methyl-ammonium cation. A new perspective on the role of A-site cation could help in improving power efficiency and accelerating the application of PSCs.
\end{abstract}

\maketitle

Solar energy is a highly efficient and eco-friendly source for future energy harvesting. In particular, inorganic/organic PSCs of ABX3 (A = Cs+, CH3NH3+, etc.; B = Pb2+; X = Cl-, Br- or I-) show extraordinary solar cell efficiencies exceeding 22 $\%$ \cite{1} with unusual characteristics,\cite{2,3,4} which attracts tremendous attention as most promising large-scale solar energy conversion materials.\cite{5} One key origin of high efficiency arises from high carrier mobility ($\mu$) $10^1-10^3$ ($cm^2V^{-1}s^{-1}$)\cite{6,7,8} even in the presence of defects.\cite{9} While the carrier transport exhibits remarkable features in experiments, there is still a gap of understanding. One aspect of this difficulty stems from significant electron-phonon (e-ph) interactions,\cite{2,8,10,11,12,13} which complicates band pictures. Another aspect is due to the A-site cation that rotates\cite{14} or even diffuses across the material, causing I-V hysteresis.\cite{15} Although recent discovery of various types of cations has greatly advanced the efficiency of PSCs,\cite{16} it also has brought about more ambiguity on the role of cations. 

In general, electrons in polar crystal experience the deformation potential in addition to Bloch potential due to large polarity of ionic bonding. The charge carriers are then described by polaron quasiparticles originating from coupling between electrons and phonons. For PSCs, there has been a critical debate on the source of e-ph coupling, acoustic vs. longitudinal. From the temperature dependence of $\mu\simT^{-3/2}$ around room temperature, acoustic phonons have been considered as the main source of e-ph coupling. \cite{11,17} However, a theoretical study of acoustic phonon scattering predicts\cite{18} $\mu$ to be several 10$^3$ $cm^2V^{-1}s^{-1}$, far above the experimental values. Meanwhile, recent studies suggest LO phonons as the main source of polaron states.\cite{19,20} The working temperature of cubic CH3NH3PbI3 (~330 K)\cite{21} is well above the Debye temperature $\Theta_D$ $\sim$175 K\cite{22} which is a representative excitation energy scale of the highest phonon mode. Therefore, it is plausible that at working temperatures of PSC, high energy PbI3 LO phonons are largely populated. 

In this study, we unveil a hidden role of A-site cation in polaron picture of PSCs by accounting for the e-ph interactions regarding both A-site cation and LO phonon scattering at the first principles level.\cite{18,19} We used the Frohlich polaron model\cite{23} which is suitable for large bandwidth ($W$) limit $W \gg \hbar \omega_{ph}$ by considering Frohlich vertex.\cite{24} We cast light on the role of various organic cations in polaronic carrier transport and propose a design principle of choosing A-site cation towards higher carrier mobility of PSCs, which is very crucial for photo-created carriers to travel to collector in solar cells. The phonon dispersion, spectral function, and carrier lifetime of lead iodide (APbI$_3$) perovskite are calculated for prospective A-site cations including existing cations A = Cs$^+$, CH$_3$NH$_3^+$ (methyl-ammonium cation; MA$^+$) based on density functional theory (DFT), density functional perturbation theory (DFPT), and many-body theory.\cite{24} We find a novel design principle for A-site cation and uncover some of unexplored A-site cations showing better charge transport property.

\begin{figure*}
\centering
\includegraphics[width=14cm]{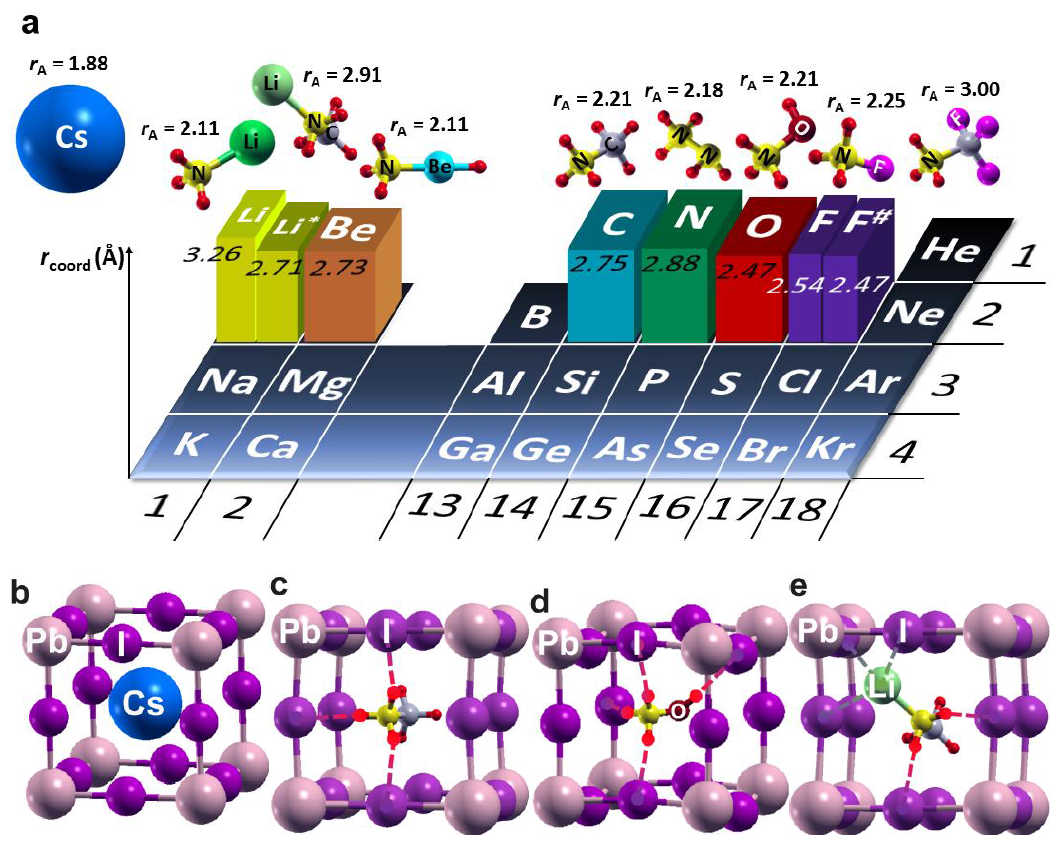}
\caption{\label{FIG. 1} a) Average values of coordinating distances rcoord to I$^-$ in cubic lead iodide PSCs for various A-site organic cations and estimated ionic radii $r_A$ of A-site cations. Li$^*$ refers to LiNH$_2$CH$_3^+$ and F$^{\#}$ refers to NH$_3$CF$_3^+$. NH$_3$BH$_2^+$ does not form molecular bonds in our DFT calculation. All units are in $\AA$. b-e) Coordination of A-site cation in cubic Pb-I cage: b) very weakly dodeca-coordinated Cs$^+$ by I$^-$, c) strongly tri-coordinated MA$^+$ by three I$^-$\textperiodcentered \textperiodcentered \textperiodcentered HN hydrogen bonds, d) strongly tetra-coordinated NH$_3$OH$^+$ (HA$^+$) by three I$^-$\textperiodcentered \textperiodcentered \textperiodcentered HN and one I$^-$ \textperiodcentered \textperiodcentered \textperiodcentered HO hydrogen bonds, e) strongly penta-coordinated LiNH$_2$CH$_3^+$ (LM$^+$) by two I$^-$\textperiodcentered \textperiodcentered \textperiodcentered HN hydrogen bonds and three I$^-$\textperiodcentered \textperiodcentered \textperiodcentered Li ionic bonds.}
\end{figure*}

There has been a little understanding regarding the role of A-site cation particularly associated with polaron nature of the material. It is important for the size of A-site cation to be suitable for a semi-empirical factor called the Goldschmidt tolerance factor $t = (r_A + r_I) / \sqrt{2}(r_{Pb} + r_I)$,\cite{25} where its ionic radius of A-site cation $r_A = r_{cm}+ r_{ion}$ with $r_{cm}$ being a distance between the center of mass and the farthest atom and $r_{ion}$ being the corresponding ionic radius except hydrogen.\cite{26} Among many candidates, inorganic cations are not appropriate according to the Goldschmidt factor because of its small ionic radius, $t < 1$. The size of A-site cation also controls a degree of rotation of octahedral, thereby tuning the bandgap of PSCs.\cite{27} One of origins of the structural instability of most-widely used MAPbI$_3$ is from its small $r_A$ being $t \sim 0.9 < 1.0$. One way to circumvent this problem is to tune the size of A-site cation (or halides), and many researchers tried to increase the number of carbon chain or amine group of MA+.\cite{26,28} As the low energy spectra near the Fermi level ($E_F$) contain only Pb$^{2+}$ (conduction) and I$^-$ (valence) states without A-site cation, the role of A-site cations has been overlooked with a limited understanding as a structural tuner with its ionic radius. 

For polar semiconductors, strong ionic bonding forms a dipole $p \sim e\omega_q^{1/2}Z^*_{\kappa}e(q)e^{iqR} $ within a unit cell, where $e$ is electron charge, $\omega_q$ is phonon frequency, $Z^*_{\kappa}$ is the Born effective charge which is a measure of polarization or dipole per unit cell, and $e(q)$ is eigenmode. Because the dipole-dipole interaction decays as $R_p^{-2}$, the interaction in momentum space diverges as $q^{-1}$ for long range components of phonon $q \to 0$.\cite{29} Only the LO mode in long range limit provides a major contribution to phonon coupling to band edge states. \textit{Ab initio} coupling matrix element of this long range part of phonon mode $\nu$ and bands n, m at wave vector $q$ is given by\cite{24}

\begin{widetext}
\begin{eqnarray}
g_{nm}(k,\omega) =
   i \frac{4\pi}{\Omega} \frac{e^2}{4\pi\epsilon_0} \sum_{\kappa,G \neq -q} \Big( \frac{\hbar}{2NM_\kappa\omega_{q\nu}} \Big)^{1/2} \frac{(q+G)Z^*_{\kappa}e_\kappa(q)}{(q+G)\epsilon_\infty(q+G)} \bra{\psi_{mk+q}} e^{i(q+G)(r-\tau_\kappa)} \ket{\psi_{nk}}
\end{eqnarray}
\end{widetext}

where $\Omega$ is the unit cell volume, $N$ the number of unit cells in the supercell, $\kappa$ atom index within a unit cell, $\epsilon_0$ the vacuum permittivity, $\hbar$ the reduced Planck constant, $\ket{\psi_{nk}}$ and $\ket{\psi_{mk+q}}$ are initial and final electronic states. In order to reduce e-ph interactions that hinder carrier transport in semiconductor, we need to decrease the dipole of a given unit cell ($p$) , i.e., Born effective charge $Z^*_{\kappa}$ and vibrational eigenmode component $e(q)$.

From a given observation in Equation (1), we note a novel design principle of A-site cation towards high efficiency of PSCs. A simple rule is to let the A site cation be nearer to I$^‒$ and to increase the number of A site cations coordinating to I$^‒$. We expect that the well-chosen cations are able to suppress e-ph coupling effectively by decreasing the $Z^*_{\kappa}$ of Pb and I with decreased distance from the A cation to I$^‒$ and by reducing   of Pb2+ and I- ions with increased coordination number of organic cations. A recently measured reorientation time scale for the MA+ molecule in a pseudo-cubic cage ($\sim 14 ps$) is long enough to participate in lattice vibrations of Pb-I frame to form the polaron state.\cite{14} Stemming from the widely used NH$_3$CH$_3$$^+$ (MA$^+$), we have substituted CH$_3$ by Li (NH$_3$Li$^+$), BeH, BH$_2$, NH$_2$ (NH$_3$NH$_2$$^+$: hydrazinium: HZ$^+$), OH (NH$_3$OH$^+$: hydroxyl-ammonium cation: HA$^+$), F (NH$_3$F$^+$: fluoro-ammonium cation, which is denoted as fA$^+$ in consideration that FA$^+$ is ofent used to denote formamidinium cation) and calculated their average coordination distance and ionic radius. While HZ$^+$ and HA$^+$\cite{30,31} are available in laboratories, fA$^+$ is difficult to be synthesized, though its spectra were observed from experiments.\cite{32} However, to understand the role of cations in the A-site, we investigated the series of cations mentioned above. The HZ$^+$ results are found to be very similar to MA$^+$, so the discussion of HZ$^+$ is omitted here to reduce the complexity in comparison, and we will focus on the comparison of Cs$^+$ with MA$^+$, HA$^+$ and fA$^+$. Using DFT calculations, we find that the average coordination distance rcoord for MA$^+$ (2.75 $\AA$) decreases to 2.47 $\AA$ for HA$^+$ and 2.54 $\AA$ for fA$^+$ (Figure 1a). These shortened ionic hydrogen bonds between I$^-$ and H(NH$_2$) in HA$^+$/fA$^+$ (decreased by 0.2-0.3 $\AA$ with respect to C) give smaller  (Pb, I) than that in MA$^+$. This suggests better performance by A-site cations with OH or F than CH$_3$. Another way to achieve our goal is fluorinating H atoms of CH$_3$ to form NH$_3$CF$_3$$^+$ (tri-fluorinated methyl ammonium: TA$^+$) or substituting H-N to Li-N to form LiNH$_2$CH$_3$$^+$ (lithium-methylamine: LM$^+$). LM$^+$ is again difficult to be synthesized, though its spectra were observed from experiments.\cite{33} This is however also investigated to understand what kind of molecular type is useful to reduce the e-ph coupling.

\begin{figure}
\centering
\includegraphics[width=8.6cm]{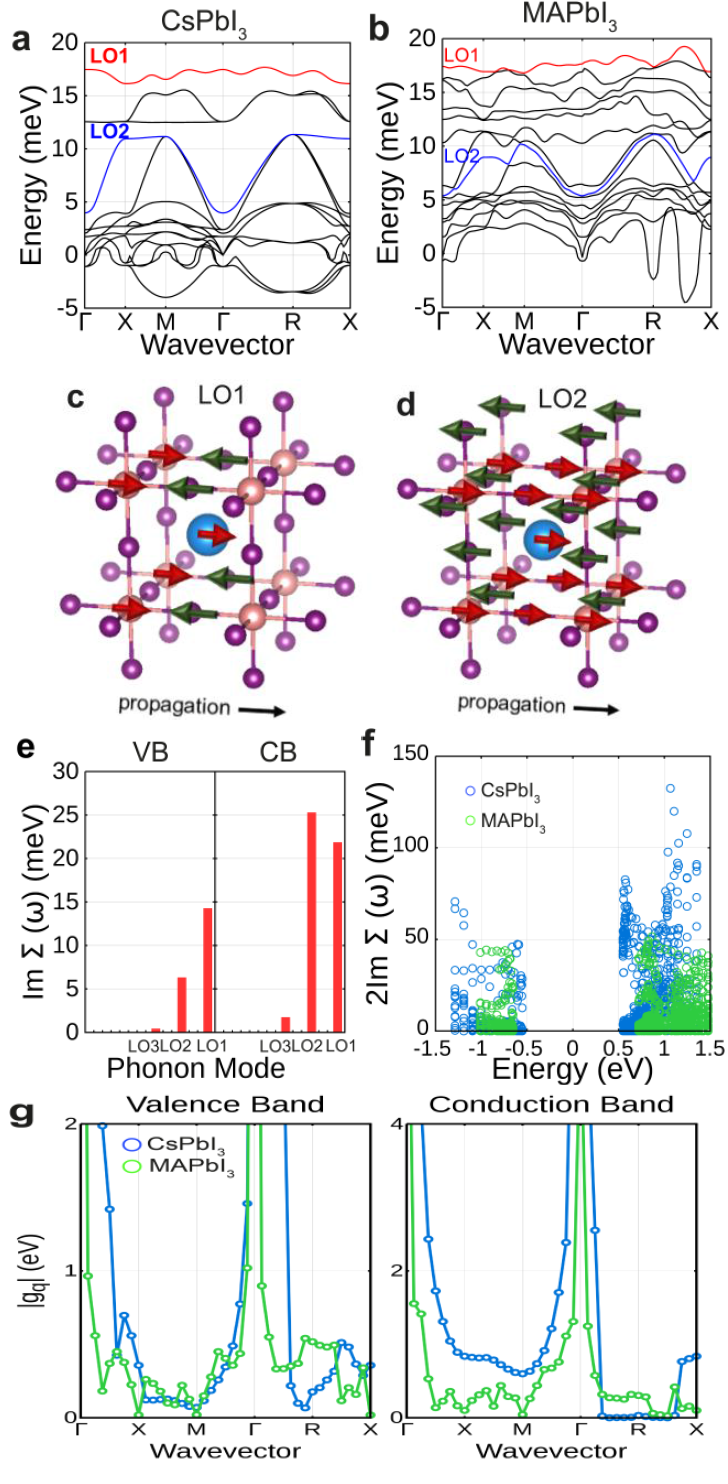}
\caption{\label{FIG. 2} a) CsPbI$_3$ and b) MAPbI$_3$ with LO1 (red) and LO2 (blue) that mostly contribute to e-ph coupling in VB and CB. The eigenmode of c) LO1 and d) LO2 in CsPbI$_3$. e) Electron-phonon self-energy of VB and CB at R(0.5, 0.5, 0.5) of BZ with three dominant LO modes: LO1, LO2, and LO3 (3rd highest LO mode).  f) Scattering rate $\tau^{-1}$ or $2Im\{\Sigma(k,\omega)\}$ for CsPbI$_3$ (blue) and MAPbI$_3$ (green) along $(\eta,\eta,\eta)$ to R(0.5, 0.5, 0.5) and R(0.5, 0.5, 0.5) to $(0.5, \eta, 0.5)$ with $0.45 \ge \eta \ge 0.5$at 330 K. $E_F$ is set to 0 eV. g) Fröhlich vertex $(m = n)$ of CsPbI$_3$ (blue) and MAPbI$_3$ (green) for VB (left)  and CB (right) with the typical divergence of polar materials for long range interactions.}
\end{figure}

Various types of configurations of A-site cations with surrounding I$^-$ ions are observed using DFT geometry optimization (Figures 1b to 1e). Inorganic Cs$^+$ is very weakly dodeca-coordinated by I$^-$ with rather large distances (~ 4.5 $\AA$). A typical A-site cation MA$^+$ forms three strong ionic hydrogen bonds between I$^-$ and H, namely tri-coordination (Figure 1c). HA$^+$ forms additional coordination employed by OH and I$^-$ (Figure 1d). Moreover, LM$^+$ having Li$^+$ instead of proton can attach noncovalently to five I$^-$s as two I$^-$H\textperiodcentered \textperiodcentered \textperiodcentered N hydrogen bonds and three I$^-$\textperiodcentered \textperiodcentered \textperiodcentered Li ionic bonds (Figure 1e). Examples of these cations show that we can actually steer the coordination number from tri- (Figure 1c), tetra- (Figure 1d) to penta-coordination (Figure 1e) with still very strong bonding strength by employing different organic cations.

Since A-site cations are not located in the low energy spectrum of electronic structure at all, they are hardly expected to participate in carrier transport. However, it turns out that the choice of A-site cation is crucial because of polaronic nature of PSCs. It is clear from the examples of two well-known materials: cubic CsPbI$_3$ and MAPbI$_3$. Phonon dispersions (Figure 2a, 2b) show a large LO(longitudinal optical)-TO(transverse optical) splitting due to the large Born effective charges of Pb and I. Long range macroscopic Coulomb interaction in non-analytic dynamical matrix of polar crystal at the $\Gamma$ point introduces the LO-TO splitting, described by the Lynddane-Sachs-Teller (LST) relation.\cite{34} Although LST does not hold exactly in polyatomic case such as PSCs, this relation can be a measure of crystal polarizability and we tabulated it in Table 1 as a reference. Organic cations including MA$^+$ diminish the splitting (Table 1), indicating that they effectively lower long range Coulomb interactions. Compared to CsPbI$_3$ (17.47 meV), the energy ($E_{LO1}$) of the highest LO mode (LO1) decreases in the case of MAPbI$_3$ (16.21 meV). In CsPbI$_3$, the clear separation between heavy Cs$^+$ modes (which mostly have low energies ~5 meV) and PbI$_3^-$ optical phonon modes (> 5 meV) is noted (Figure 2a). In MAPbI$_3$, the rotation/translation modes of MA+ are mixed with the PbI$_3^-$ LO modes (Figure 2b). A light molecule such as MA$^+$ which interacts with I$^-$ by anisotropic directional hydrogen bonding breaks the symmetry of PbI$_3$  lattice, and effectively suppresses the PbI$_3^-$ LO lattice vibrations by the linear I$^-$\textperiodcentered \textperiodcentered \textperiodcentered HN halogen-hydrogen bonding perpendicular to the Pb$^{2+}$-I$^-$-Pb$^{2+}$ perovskite skeleton. 

A calculation of e-ph coupling for all modes reveals that most contributions of e-ph scattering in both CsPbI$_3$ and MAPbI$_3$ at band edges of valence band (VB) and conduction band (CB) come from LO modes: the highest LO mode LO1 (red) and the second highest LO mode LO2 (blue) (Figure 2a, 2b). Except LO1 (Figure 2c) and LO2 (Figure 2d), other 13 modes' contributions are almost negligible or completely insignificant (Figure 2e). Therefore, by observing these two dominant modes, it would be easy to track the changes in e-ph coupling, once other cations are substituted for the MA$^+$ in the A-site. In CsPbI$_3$, due to large polarizability of the Pb$^{2+}$-I$^-$ bond and large proportion of PbI$_3^-$ LO vibration, the e-ph scattering is much larger than in MAPbI$_3$ (Figure 2f). As the inorganic cation Cs$^+$ is replaced by MA$^+$, the e-ph coupling diminishes because of the suppression of PbI$_3^-$ LO phonon modes by MA$^+$ which is attached to I$^-$ atoms. The result agrees with our design principle that organic cations like MA$^+$ should decrease the scattering rate compared to inorganic PSCs (Figure 2f). The Fröhlich vertex  $(m = n)$ of CsPbI$_3$ (blue) and MAPbI$_3$ (green) for LO1 and LO2 indicates that in addition to the overall reduced e-ph couplings, long range interactions ($q \to 0$) are substantially quenched for MAPbI$_3$ (Figure 2g). These overall observations point out that properly chosen organic cations behave much better in charge mobility than inorganic ones.

Negative (imaginary) acoustic modes, for example at M (tetragonal) and R (orthorhombic) in CsPbI$_3$, indicate instability of cubic structure at 0 K. However, because we are solely interested in the optical modes, particularly LO, our calculation can be justified if LO spectra do not change much at high temperatures. Fourier transformed velocity auto-correlation function (VAC) from ab initio molecular dynamics (MD) simualtions at ~ 600 K above the critical temperature $T_C$ (~ 583 K)\cite{35} for the cubic phase (Figure S1a, Supporting Information) shows that the overall MD spectra agree with 0 K lattice phonon density of states (DOS), and both LO1 and LO2 energy DOS do not change much (Figure S1b, Supporting Information). Therefore, we assume that at high temperatures, phonon spectra of high energy LO modes are not severely affected.

Among the tested A-site cations, we have chosen the most promising cations: HA$^+$, fA$^+$, LM$^+$, and TA$^+$ according to the coordination number [MA$^+$(3) $<$ HA$^+$(4) $<$ LM$^+$(5)] and the coordinating distances [d(fA$^+$), d(TA$^+$), d(HA$^+$) $<$ d(MA$^+$)]. The Born effective charges of I atoms coordinated by those organic cations decrease as   $ Z^{*}_{MAPbI_3} > Z^{*}_{LMPbI_3} > Z^{*}_{HAPbI_3} > Z^{*}_{fAPbI_3} > Z^{*}_{TAPbI_3}$ (Table 1). Being substituted by O, F, or being fluorinated, the cations working as a buffer are able to reduce the effective Pb-I dipole. The coordination of cations can reduce $Z^*$ up to 40 $\%$ compared to non-coordinated one in the case of fAPbI3 (table S1, Supporting Information). 

The coordination number is also important since Pb$^{2+}$-I$^{-}$ LO eigenmodes are absorbed by organic cations via coordination. (Figure 3) The increase in coordination number decreases the value of  , for example, HA$^+$(1.26) (Figure 3b) and LM$^+$(1.33) (Figure 3d) compared to MA$^+$(1.46) (Figure 2b and Table 1). LMPbI$_3$ shows a helpful example of how the Pb-I motion loses its contribution in LO1 and LO2 as the number of organic cations coordinated to I increases. As for the LM$^+$ cation (Figure 3e), the degree of mixing and screening of LO modes by the cation is maximized among the candidate cations by halogen-hydrogen bonding and directional linear I$^-$\textperiodcentered \textperiodcentered \textperiodcentered LiNH$_2$ halogen-lithium bonding, followed by HA$^+$. The HA$^+$ cation being tetra-coordinated to I also absorbs the Pb-I LO mode more significantly than the MA$^+$ cation. (Figure 3e)

\begin{figure}
\centering
\includegraphics[width=8.6cm]{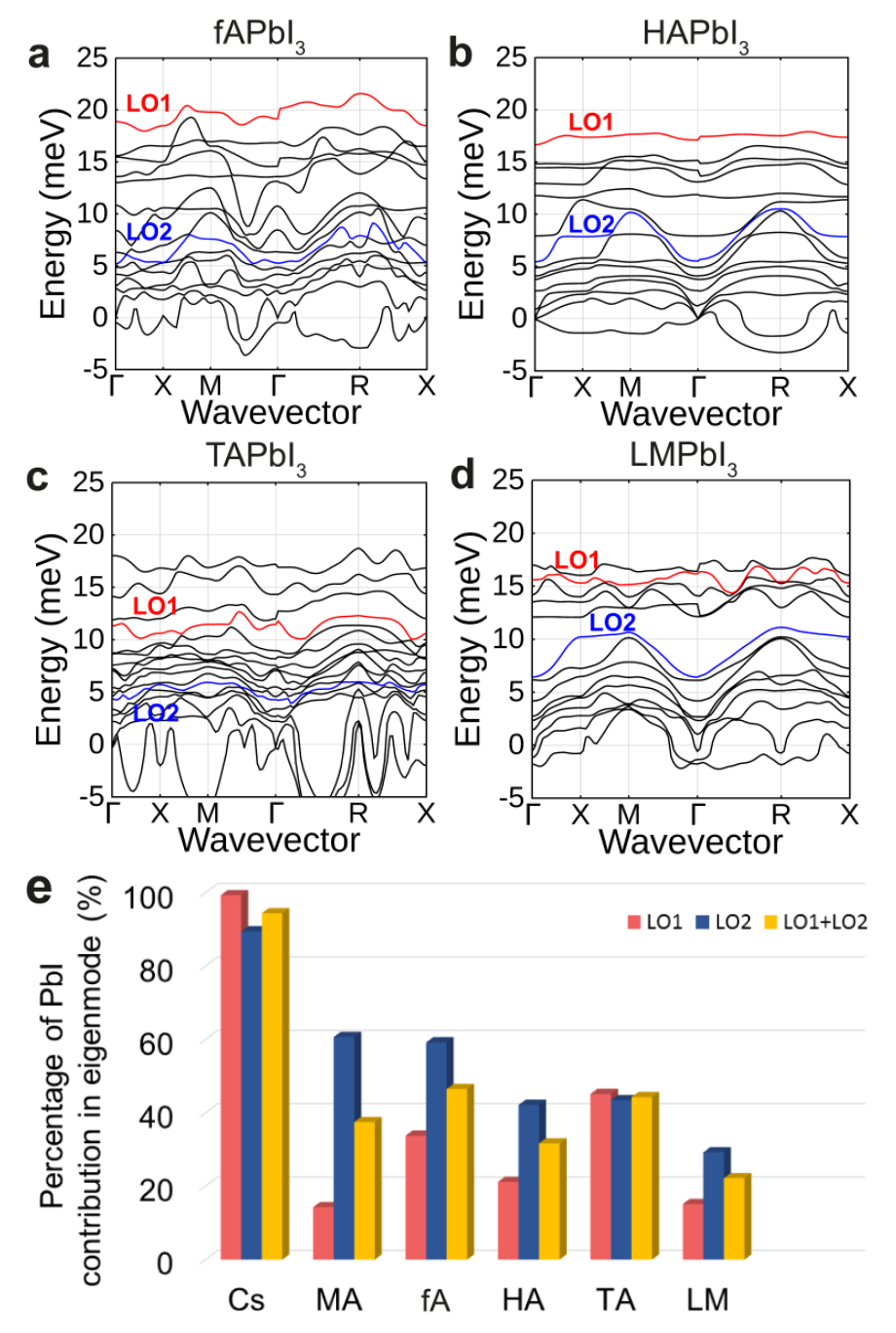}
\caption{\label{FIG. 2} Vibrational dispersions of a) fAPbI$_3$, b) HAPbI$_3$, c) TAPbI$_3$, d) LMPbI$_3$. e) Histogram of dynamical matrix components of LO eigenmodes for LO1 (red), LO2 (blue), and its average (yellow). A small proportion of A-site cation component in LO modes in CsPbI3 substantially increases by modes of the organic cations. Eigenmode contributions of A-site cation in LO modes are Cs$^+$ (5.6 \%), MA$^+$ (62.5 \%), fA$^+$ (53.5 \%), HA$^+$ (68.3 \%), TA$^+$ (55.7 \%), and LM$^+$ (77.8 \%). The eigenmode components of Pb-I are absorbed by A-site organic cations. }
\end{figure}

The overall discussions can be summed up in Figure 4. It turns out that, in VB and CB, TA$^+$ which was once conjectured as a promising cation is not a good candidate for carrier transport, because a new LO motion between F and Pb-I is present in TAPbI$_3$; this additional LO mode gives rise to the largest e-ph scattering (Figure S2, Supporting Information). In VB, the scattering rates of HA$^+$ and LM$^+$ are affixed to the value of MA$^+$. However, it is interesting to note that the fA$^+$ cation, which buffers a large polarizability between Pb and I (Table 1), reduces the scattering rate to half the value of MA$^+$ (Figure 4a). In CB, a reduction of the scattering rate is clear for fA$^+$, HA$^+$, and LM$^+$. In particular, LM$^+$ significantly reduces the e-ph scattering (Figure 4b) because of its largest mixing with Pb-I LO modes (Figure 3e). Electron self-energy of Rayleigh-Schrodinger perturbation theory for nondegenerate conduction band minimum at $\Gamma$ of Brillouin Zone (BZ) is simply given by $Re \{ \Sigma(k,\omega) \} = -\alpha \hbar \omega_{LO}$ at 0 K, where $\alpha$ is polaron coupling constant, $\alpha = \frac{e^2}{8\pi\epsilon_0\hbar\omega_{LO}} \Big( \frac{2\omega_{LO}m_0}{\hbar} \Big)^{1/2} \Big(\frac{1}{\epsilon_\infty} - \frac{1}{\epsilon_0} \Big)$.\cite
{36} Although this relation is from a simple model Hamiltonian, still  can provide a good measure of polaron coupling for real materials such as PSCs. Thus, the comparison of polaron coupling constants at 0 K [$\alpha$(CsPbI$_3$) = 1.42, $\alpha$(MAPbI$_3$) = 0.11, $\alpha$(HAPbI$_3$) = 0.036, and $\alpha$(LMPbI$_3$) = 0.026 (Figure 4c)] reveals that the prospective organic cations reduce the polaron coupling effectively. As confirmed by other DFT calculations,\cite{37} we note that strong hydrogen bonding of cations introduces a change of simple quadratic band minimum (or maximum) into a double well like shape (Figure 4c). Photo-luminescence (PL) bandwidths as the function of temperatures (Figure 4d) confirm that HA$^+$ and fA$^+$ with the strong coordination to I and LM$^+$ with a large coordination number to I are promising A-site organic cations for highly efficient PSCs. Finally, from a concise relation for the mobility $ \mu = e\tau/m^* (cm^2V^{-1}s^{-1})$, where $e$ is carrier charge, $\tau$ is the calculated lifetime of PSCs, and $m^*$ the effective mass of CB and VB,\cite{38} we estimate the mobility for the prospective materials at 330 K for electrons, $\mu_{TAPbI_3}(61.85) < \mu_{CsPbI_3}(73.94) < \mu_{MAPbI_3}(118.08) < \mu_{HAPbI_3}(150.82) < \mu_{fAPbI_3}(162.92) < \mu_{LMPbI_3}(464.41)$, and for holes $\mu_{TAPbI_3}(57.88) < \mu_{CsPbI_3}(98.11) < \mu_{HAPbI_3}(102.22) < \mu_{MAPbI_3}(108.70) < \mu_{LMPbI_3}(110.25) < \mu_{fAPbI_3}(226.77)$.

\begin{figure}
\centering
\includegraphics[width=8.6cm]{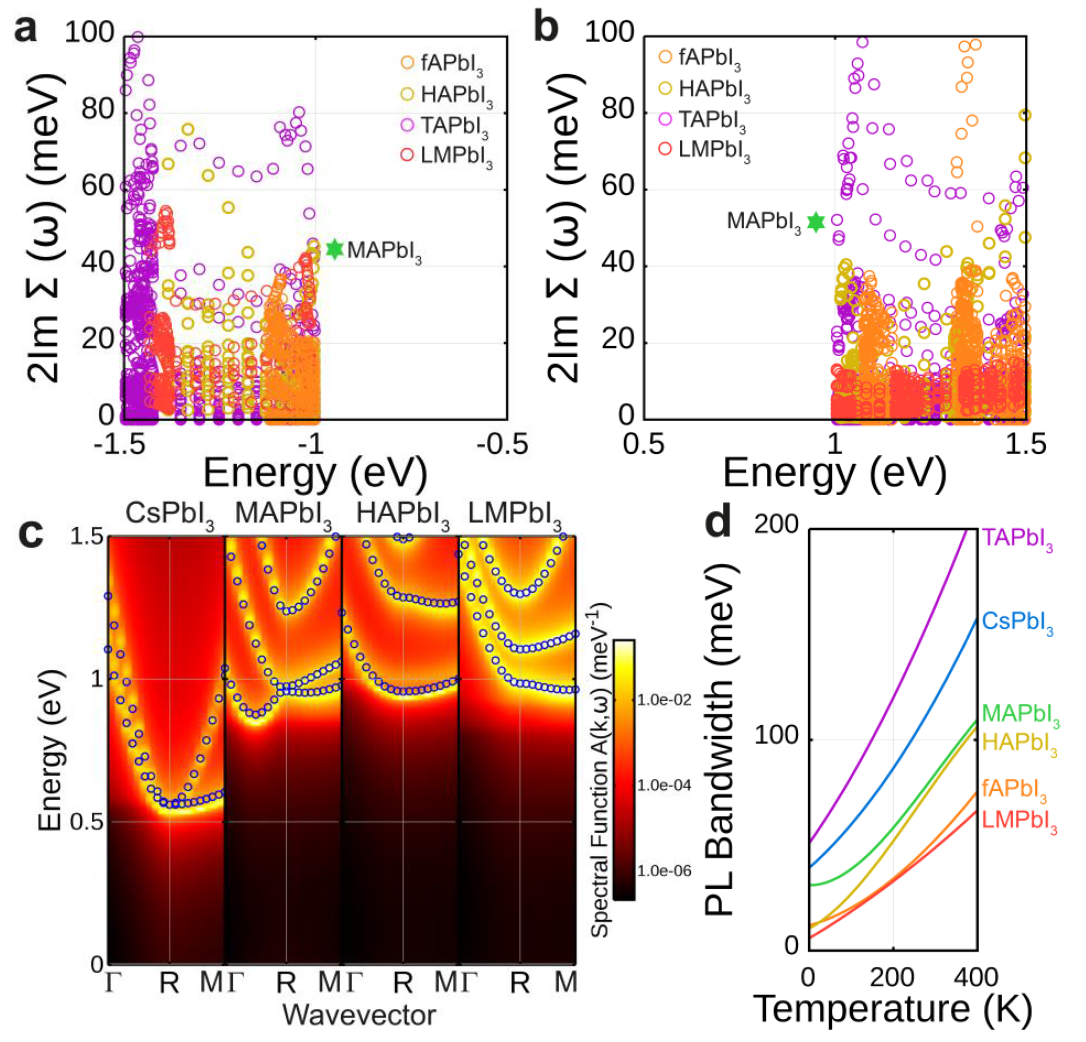}
\caption{\label{FIG. 2} a) Hole and b) electron scattering rate $\tau^{-1}$ or $2Im\{\Sigma(k,\omega)\}$ for fAPbI$_3$(orange), HAPbI$_3$(yellow), TAPbI$_3$(purple), LMPbI$_3$(red) at band minimum or maximum along k = $(\eta,\eta,\eta)$ to R(0.5, 0.5, 0.5) and R(0.5, 0.5, 0.5) to $(0.5,\eta,0.5)$ with $0.45 \ge \eta \ge 0.5$at 330 K. Band maxima(minima) are aligned so that it can be easily recognized. of $2Im\{\Sigma(k,\omega)\}$ MAPbI$_3$ at band edges are indicated as star (green). c) Electron spectral function of CsPbI$_3$, MAPbI$_3$, HAPbI$_3$, and LMPbI$_3$ along with DFT Kohn-Sham eigenvalues (blue circles) in the vicinity R(0.5, 0.5, 0.5) of BZ at 0 K. Real part of electron self-energy $Re\{\Sigma(k,\omega)\}=\alpha\hbar\omega_{LO}$ indicates the strength of e-ph coupling for various A-site organic cations. d) Photo-luminescence (PL) bandwidth as the function of temperatures for TAPbI$_3$(purple), CsPbI$_3$(blue), MAPbI$_3$(green), HAPbI$_3$(yellow), fAPbI$_3$(orange), and LMPbI$_3$(red).}
\end{figure}

A-site cations have not been considered as a main factor for carrier transport but been neglected except its role in structural tolerance for cubic perovskite. Here, we shed a light on the hidden role of A-site cation in PSCs within Fröhlich polaron picture. From this perspective, we found a design principle of choosing a suitable cation for a better carrier mobility to boost efficiency of PSCs. The key for steering the charge transport is the coordination of organic cations to I atoms. By increasing the interaction strength of the A-site cation to I and the number of organic cations coordinated to I, the vibrational mode of molecular cations involved in directional I-…HN(O) halogen-hydrogen or I$^-$\textperiodcentered \textperiodcentered \textperiodcentered Li bonding weakens e-ph interaction because of the reduction in $Z^*(Pb, I)$ and eigenmode components of Pb and I. We have calculated polaron properties of lead iodide PSCs with various types of A-site cations from first principles. As compared with the most-used MA$^+$ organic cation and inorganic cation Cs$^+$, the untapped organic cation HA$^+$ and other cations similar to LM$^+$ are likely to be highly promising for light harvesting. Because most of hole conducting materials (HTM) contain Li$^+$ ions in lithium-bis(trifluoromethylsulphonyl)imide (LiTFSI), the formation of LM$^+$ at the interface of HTM could be possible by diffusion of Li ions followed by heating in consideration that LM$^+$ is not necessary to be one molecular cation but a molecular system of Li$^+$\textperiodcentered \textperiodcentered \textperiodcentered NH$_2$CH$_3$ where a Li$^+$ cation is solvated by NH$_2$CH$_3$ and three I$^-$ anions. This is quite plausible in the presence of three I$^-$ anions because Li$^+$ is known to be solvated by NH$_2$CH$_3$,\cite{39,40} and hence it is considered as a possible process to provide such a system. To conclude, we have uncovered a new perspective on the role of the A-site organic cation that reduces the Frohlich e-ph coupling of the PbI$_3^-$ LO vibration. The proposed design principle could help find a new prospect of A-site organic cation for the development of PSCs to expedite effective light harvesting.

We calculated ionic radius of organic cations using PBE functional and aug-cc-pVTZ basis set. Quantum ESPRESSO package v.6.1\cite{42} were used for DFT, DFPT, and ab initio MD. We used norm-conserving PBE pseudopotential including d core states of Pb and sampled an unshifted $(8 \times 8 \times 8)$ grid in BZ for geometry optimization and for electronic calculation at energy cutoff of 80 Ry for PSCs containing organic cations and 100 Ry for CsPbI$_3$. Because lattice distortion is not significantly affected by types of functional and spin-orbit coupling,\cite{43} we used PBE exchange functional. We used DFPT for phonon dispersions and considered the non-analytic term of dynamical matrix at $\Gamma$ of BZ to account for LO-TO splitting. The e-ph coupling and electron self-energy were calculated using EPW v.4.1.0\cite{44} with unshifted $(64 \times 64 \times 64)$ q-points and with Gaussian broadening 10 meV for BZ integration. We performed ab initio MD simulations with time step t = 0.50 fs and total duration of ~10 ps at 600 K. Fourier transformed velocity auto-correlation(VAC) was done for 9 ps excluding 1 ps of initial thermalization and was smoothened by Savitzky-Golay filter.

\begin{acknowledgments}
This work was supported by National Honor Scientist Program (2010-0020414) of NRF. Computation was supported by KISTI (KSC-2016-C3-0074). C.W.M. performed all calculations. J. H. Y. calculated VAC spectrum of ab initio MD. All authors analyzed the data and wrote the manuscript.
\end{acknowledgments}

\bibliographystyle{apsrev4-1}
\bibliography{myungbib}

\end{document}